# "HOW DO YOU FEEL?": EMOTIONS EXHIBITED WHILE PLAYING COMPUTER GAMES AND THEIR RELATIONSHIP WITH GAMING BEHAVIORS

Rex P. Bringula,
College of Computer Studies and Systems, University of the East, Philippines
rex_bringula@yahoo.com

Kristian Paul M. Lugtu,
College of Computer Studies and Systems, University of the East, Philippines
contact@kristianlugtu.com

Mark Anthony D. Uy,
College of Computer Studies and Systems, University of the East, Philippines
uy.markanthony@gmail.com

Ariel D.V. Aviles,
College of Computer Studies and Systems, University of the East, Philippines
co113@yahoo.com

## ABSTRACT

This descriptive study utilized a validated questionnaire to determine the emotions exhibited by computer gamers in cyber cafés. It was revealed that most of the gamers were young, male, single, as well as high school and vocational students who belonged to middle-income families. Most of them had computer access at home but only a few had Internet access at home. Gamers tended to play games in cyber cafés at least three times a week, usually in the evening, for at least two hours per visit. They also reported that they played games frequently. Majority of the gamers were fond of playing DOTA, League of Legends, and CABAL and they had been playing games for at least two years. It was disclosed that they exhibited both positive and negative emotions while playing games. It was shown that gamers were inclined to be more anxious to be defeated in a game as gaming became frequent and length of years in playing games increased. They also had the tendency to become more stressed when length of years of playing games increased. On the other hand, other gaming behaviors were not significantly related to other emotions. Thus, the null hypothesis stating that gaming behaviors of the respondents are not significantly related to the emotions exhibited while playing the computer games is partially rejected. Therefore, not all emotions exhibited while playing computer games could be attributed to their gaming behaviors. It is recommended that other emotions such as anger, frustration, boredom, amusement, etc. be included in future research.

## 1. INTRODUCTION

The need for inexpensive Internet and computer access in a Third World country like the Philippines fueled the proliferation of computer shops [18]. Computer shops are also called Internet shops, Internet cafés, or cyber cafés. This form of computer and Internet public access open venues for diverse services such as chatting, typing, printing, scanning, and gaming. Among these services, most gamers avail themselves of the gaming service [23,24]. This is also the case in the Philippines where the costs of personal computers and game software packages are still barriers.

Computer games are the favorite pastime of the youth nowadays [1, 15, 24]. They draw the attention of researchers to investigate the impact of computer games on their players [e.g.,7, 9, 16]. However, the existing literature lacks identification of different emotions exhibited while playing computer games in a competitive environment such as the Internet café. Moreover, most of the literature [e.g., 6, 8, 16] were focused on a single dimension of emotions while playing computer games. Also, previous investigations [e.g., 19] assumed that gamers had prior emotion (e.g., boredom) that triggered gamers to play computer games. It can also be argued that games could serve as stimuli to experience different emotions. Furthermore, it is also unknown if players' emotions exhibited while playing computer games are related to their gaming behaviors. The study attempted to address these research gaps. Toward this goal, it attempted to answer the following questions.
1. What is the profile of the respondents in terms of age, gender, civil status, highest educational attainment, employment, computer access at home, Internet access at home, and family monthly income?
2. What are the gaming behaviors of the respondents in terms of frequency of visit in a week, length of years playing games, games being played, hours spent per visit, and frequency of gaming?
3. What are the emotions exhibited while playing computer games?
4. Is there a significant relationship between emotions exhibited while playing computer games and the gaming behaviors of the respondents?

## 2. Literature Review
### 2.1 Theories of Emotion
Plutchik [21, p. 551] defined emotion as "an inferred complex sequence of reactions to a stimulus including cognitive evaluations, subjective changes, autonomic and neural arousal, impulses to action, and behavior designed to have an effect upon the stimulus that initiated the complex sequences." It could also be defined as "psychological states that comprise thoughts and feelings, physiological changes, expressive behaviors, and inclinations to act" [2, p. 285]. Cannon-Bard theory of emotion [12] offered an explanation on how emotions were activated. According to this theory, there were stimuli or events that could elicit emotion and eventually trigger physiological arousal and a person could experience emotions simultaneously and independently.

Emotions and how they could be measured were also explained through theories. The evaluative space model of Cacioppo et al. [10] proposed that evaluation of emotions as "good" or "bad" should be independent of each other, that is, emotions can be good and bad at the same time. This was explained further in Watson and Tellegen's [3] emotion model. In this model, it allows the possibility that positive and negative affects be independent of one another. In other words, positive and negative affects can be felt by a person at the same time.

Positive and negative affects are often referred to as the Big Two emotions [20]. The former refers to all high energy emotions that feel good or pleasurable such as energetic, engagement, joy, happiness, love, and enthusiasm [4, 20]. On the other hand, the latter refers to the unpleasant feelings such as anxiety, fear, hate, worry, distress, anger, hostility, and disgust [4, 20].

## 2.1 Different Studies Conducted on Emotions and Gaming

Hutton and Sundar [5] determined the role of emotions in helping the youth reach their creative potentials. The study defined creativity as "ideation, measured as the amount of novel ideas generated, rather than on the whole creative process, which can include problem solving, generating ideas and evaluating them" [13, p. 295]. In order to find answer to this research problem, they utilized a video game DanceDance Revolution. By varying the levels of arousal (the extent to which an individual feels energized physically or mentally) associated with low, medium, and high levels of exertion in the video game and by inducing a positive or negative mood (i.e., valence), it was found out that the emotion significantly affected creativity through the interaction of arousal and valence. Lower arousal levels resulted in higher creativity scores when coupled with a negative mood. On the other hand, a positive mood at high arousal resulted in greater creative potential than a negative mood.

The relationship of video game playing to psychological well-being, aggressive behaviors, and adolescents' perceived threat of video computer game playing was investigated in the Islamic Republic of Iran [8]. The researchers administered questionnaires which were distributed at random to 444 eight middle school sample participants. It was disclosed that participants spent an average of 6.3 hours per week playing video games and 47% of them had played one or more games with violent content. Non-gamers suffered poorer mental health compared to excessive gamers. Participants who started gaming at a younger age were more likely to score poorer in mental health. Aggressive behaviors were associated with length of gaming, and boys who played games excessively had the tendency to show more aggressive behaviors than girls. It was also revealed that those who perceived less serious side effects of video gaming and those who had personal computers were more likely to play video games excessively.

The study of Devilly et al. [6] determined whether or not gamers could change emotions during longer lengths of game play. Ninety-eight participants were randomly assigned to play the game Quake III Arena for either 20 or 60 minutes. It was shown that a change in scores for gamers indicated that short gaming led to a higher increase in anger ratings than long gaming. Their results also disclosed that female gamers showed a larger change in state anger (CSA) than the male participants. Also, those who were unexposed to video games with violent content had higher CSA than those who were exposed to these games.

Poels et al. [11] conducted an experiment to investigate how player emotions (measured through self-report and physiological recordings) during game-play could predict playing time and game preferences. Nineteen participants aged between 18 and 42 were involved in the study. The researchers used four personal computer games – two first person shooter games and two race games. Participants played each of the four games in 10 minutes. Afterwards, they rated their experiences using a Self-Assessment Manikin Scale. Participants were again requested to play the game in another laboratory for another 30 minutes. Participants were also connected to TMSi Mobi 6 Bluetooth device to determine their physiological changes. Skin Conductance Level (an indicator of arousal) and zygomaticus major and currogator supercilii electromyography (EMG) measures (indicators of (dis)pleasure) were gathered from the sensors. Data from the self-report and physiological recordings showed that pleasure was more predictive for short-term game-play (immediately after game-play). On the other hand, arousal, particularly for game preferences, was most predictive for long-term (after 3 weeks) game-play.

Lastly, Demirok et al. [16] examined the relationship of time spent playing computer games with violent content and self-reported anger of 400 students in North Cyprus. They revealed that 43% of the students played computer games 3-4 times a week while many (32%) played every day. It was shown that students who played computer games for 2 to 3 hours a day scored higher on expressed anger than those who played for less than half an hour a day. Moreover, young people who preferred action, adventure, fight, and strategy games were found to have higher levels of anger than those who played other types of computer games.

## 3. METHODOLOGY

The study employed an exploratory-descriptive design in which a descriptive-survey form was used as the research instrument. The respondents of the study were Internet users in a cyber café in Guagua, one of the municipalities of Pampanga, Philippines. It has a population of 111,199 [17] that served as basis in computing the minimum sample size. It was chosen as the research locale of the study because it is a middle-class municipality in a highly urbanized province closest to Manila. As such, computer shops were present in the locale and highly accessible. Using Sloven's formula (e = 0.10), a 100-minimum sample size was utilized. One hundred survey forms were distributed evenly in the morning (between 7 a.m. to 11:59 a.m.), afternoon (between 1 p.m. to 6 p.m.), and evening (between 6:01 p.m. to 9 p.m.; the evening session exceeded one survey form). All forms were retrieved. Respondents (regardless of age, sex, religious affiliation, etc.) answered the questionnaire.

The study adopted Plutchik's [21] and Manstead's [2] definition of emotion. Moreover, it is guided by the evaluative space model of Cacioppo et al. [10], emotion model proposed by Watson and Tellegen [3], and by the Cannon-Bard Theory. Through these theories, the survey instrument was developed. Self-reports of subjective feelings [14, 22] was adopted in the determination of gamers' emotions.

A survey form was used to gather information on the profile of the respondents such as place of residence, age, sex, occupation, religious affiliation, civil status, computer ownership, monthly family income, and educational background. This profile was included in the first part of the questionnaire. The second part gathered the gaming behaviors of the respondents in a cyber café usage in terms of hours spent per visit, visiting time in a day, frequency of visit in a week, length of years playing games, most played games, and frequency of gaming. The last part of the survey instrument allowed gamers to self-report the emotions they felt while playing games such as happiness, stress, excitement, disappointment, irritation, delight, anxiety, and pressure. The study employed a semantic differential in rating the emotions of the respondents.

The statistical tools used in the treatment of data included frequency counts, percentages, mean, and Pearson correlation. Frequency counts, percentages, and mean were utilized to describe the data. Pearson correlation was also used to determine the relationship of gaming behaviors and emotions exhibited of the respondents. In this connection, 1% level of probability with 99% reliability was followed to determine the degree of significance of the findings.

## 4. RESULTS AND DISCUSSION

Table 1 shows the profile of the respondents. Most gamers were at least twenty years old (f = 68, 68%) and were college students (f = 64, 64%). Majority of the respondents were graduating students. This was also confirmed by the employment status of the respondents (Students, f = 48, 48%). Consistent with the literature, majority of the respondents were male (f = 91, 91%) and single (f = 96, 96%) [18].

**Table 1**. Profile of the Respondents

| Profile | *f* | % |
|---|---|---|
| **Age*** | | |
| 19 and below | 32 | 32 |
| 20 and above | 68 | 68 |
| *Average age is 21 years old.* | | |
| **Gender** | | |
| Male | 91 | 91 |
| Female | 9 | 9 |
| *Civil Status* | | |
| Single | 96 | 96 |
| Married | 4 | 4 |
| **Highest Educational Attainment** | | |
| Elementary | 2 | 2 |
| High school | 34 | 34 |
| College / Vocational | 64 | 64 |
| *Employment* | | |
| Student | 48 | 48 |
| Employed | 34 | 34 |
| Not employed | 18 | 18 |
| *Computer Access at Home* | | |
| With computer access at home | 79 | 79 |
| Without computer access at home | 21 | 21 |
| **Internet Access At Home** | | |
| With internet access at home | 19 | 19 |
| Without internet access at home | 81 | 81 |
| **Family Monthly Income** | | |
| Lower income class (less than P14,000 / $325) | 1 | 1 |
| Middle income class (Php 14,000 – Php 56,000 / $326 to $1300) | 99 | 99 |
| **TOTAL** | **100** | **100** |

It is interesting to note that majority of the respondents had computer access at home (f = 79, 79%). However, Internet access was still a barrier (f = 81, 81%). The affordability of computers due to their decreasing cost enables a person to own a personal computer. However, this is not the case for Internet access. Internet access is still relatively expensive. A monthly Internet subscription fee of Php1,000 (US$ 25) is not seen as a primary household commodity. Further, such fee is relatively expensive for a family who belongs to a middle-income class (f = 99, 99%).

Meanwhile, Table 2 shows the gaming behavior of the respondents in terms of frequency of visit in a week, visiting time in a day, length of years of playing games, most played games, hours spent per visit, and a self-report frequency of gaming. The findings in terms of frequency of visit in a week revealed that about 52% (f = 52) visited cyber cafés at most four times a week while there were gamers visited cyber cafés for at least five times a week (f = 48, 48%). This finding shows that there was almost an even distribution of occasional and frequent visitors of cyber cafés to play computer games. While others tended to visit cyber cafés to play games prudently, others exhibited a more recurrent behavior of visit. Clearly, the disparity could not be attributed to the profile of the respondents since the profile of the respondents was homogenous. It can be deduced that other factors could explain this phenomenon such as self-restraint, self-discipline, or influence of other people (e.g., parents that monitor their children).

**Table 2.** Gaming Behavior of the Respondents

| Gaming Behavior | *f* | *%* |
|---|---|---|
| **Frequency of Visit in a Week** | | |
| Once to twice a week | 14 | 14 |
| Three to four times a week | 38 | 38 |
| Five to six times a week | 42 | 42 |
| Every day | 6 | 6 |
| **Visiting Time in a Day** | | |
| Morning | 0 | 0 |
| Afternoon | 39 | 39 |
| Evening | 61 | 61 |
| **Length of Years Playing Games** | | |
| Less than 1/2 year | 0 | 0 |
| 1/2 year to less than 1 year | 6 | 6 |
| 1 year to less than 1.5 years | 11 | 11 |
| 1.5 years to less than 2 years | 4 | 4 |
| 2 years to less than 2.5 years | 39 | 39 |
| 2.5 years or more | 40 | 40 |
| **Three Most Played Games** | | |
| DOTA | 86 | 86 |
| League of Legends | 66 | 66 |
| CABAL | 59 | 59 |
| **Hours Spent per Visit** | mean = 2.73 hours | |
| **Frequency of Gaming** | mean = 4.06 (Frequent) | |

Gamers spent more than 2 hours (mean = 2.73 hours) per visit in playing games. This finding gives a clear gaming behavior of the respondents in terms of time expenditures. It can be inferred that the respondents of this study could prudently spend their time in playing games. In other words, they tended to visit cyber cafés in a relatively short but frequent manner. A self-report frequency of visit confirmed this conclusion (mean = 4.06, frequent).

Gaming in cyber cafés was also done predominantly during the night (f = 61, 61%). This is not surprising since gamers had more time to play during the night when their classes were over than during the rest of the day when they had to go to school for their studies.

Table 3 shows the emotions exhibited by gamers. There were three positive and five negative emotions exhibited while playing games. Gamers reported they were very happy (mean = 4.56) when they were playing computer games. Excitement (mean = 4.22) and delight (mean = 4.33) were also dominant emotions. These positive emotions signify that gamers showed cheerful feelings when they were playing computers games.

Interestingly, they also felt negative emotions while playing computer games. They felt anxiety (mean = 3.90), irritation (mean = 4.00), and stress (mean = 4.15). On the other hand, they felt pressure on a lesser extent when they were competing with a better player. Stress was the most prominent negative emotion felt by computer gamers. These findings confirm the assumptions of the study that mixed emotions were shown while playing computer games.

This can be explained by the nature of the games they play. The top three most played games require gamers to dedicate time to establish game dominance in terms of power, game incentives, game character's life and strength, and watchtowers. In gamers' point of view, these will be of waste when they are defeated in a game. Hence, maintaining game dominance becomes stressful. This is confirmed in Table 4.

**Table 3**. Emotions Exhibited While Playing Computer Games

| Question (*Emotions*) | Mean | V. I. | Type of Emotion |
|---|---|---|---|
| How happy are you when you play computer games? (*Happiness*) | 4.56 | Very happy | + |
| What is your stress level when you are playing computer games? (*Stress*) | 4.15 | Stressed | – |
| How excited are you when you play computer games? (*Excitement*) | 4.22 | Excited | + |
| How disappointed are you when you lose in a game? (*Disappointment*) | 3.95 | Disappointed | – |
| How irritated are you when you are interrupted in your game? (*Irritation*) | 4.00 | Irritated | – |
| How delighted are you when you win a game? (*Delight*) | 4.33 | Delighted | + |
| How anxious are you when you are defeated in a game? (*Anxious*) | 3.90 | Anxious | – |
| How pressured are you when your playmate is better than you? (*Pressure*) | 3.43 | Moderately pressured | – |

As shown in Table 4, length of playing computer games was significantly related stress ($r = 0.33$) and anxiety ($r = 0.36$). The results are unlikely to have arisen from sampling error ($p < 0.01$). The findings indicate that stress level and the concern to be defeated in a game build up when gamers' length of years playing games increases. As the game level of the player progresses, it becomes more difficult for a gamer to achieve a higher level. Consequently, the competitive environment becomes stiffer. This induces more stress to the gamers.

Meanwhile, frequency of gaming was also related to anxiety ($r = 0.27$, $p < 0.01$). This can be explained by the fact that games have to be constantly played so that the game characters will be more dominant and powerful. Gamers need to invest effort, time, and money not only to achieve a remarkable game status but also to retain such status. Thus, a single mistake in game strategy may lead to game defeat. In turn, game defeat may lead to anxiety.

It was observed that anxiety had significant and positive relationship to two gaming behaviors (i.e., frequency of gaming and length of years playing games). It implies that anxiety was more sensitive to gaming behaviors. In other words, it would take at least one gaming behavior to trigger gamers' anxiety towards the game while stress influenced only one gaming behavior (i.e., length of years playing games).

It is also interesting to note that emotions related to gaming behaviors were negative emotions. None of the positive emotions had significant relationship to gaming behaviors. This implies the possibility that higher game engagement may lead to the gamers' expression or manifestation of negative emotions. Nonetheless, the study emphasized that these negative emotions were exhibited *only* during the game. The study did not establish that these emotions were carried over in the gamers' actual lives.

**Table 4**. Relationship between Emotions Exhibited while Playing Games and Gaming Behavior

| Emotions | Gaming Behaviors[a] | | |
|---|---|---|---|
| | Hours Spent per Visit | Frequency of Gaming | Length of Years Playing Games |
| Happiness | $r = 0.16$, $p = 0.101$ | $r = 0.11$, $p = 0.260$ | $r = 0.04$, $p = 0.708$ |
| Stress | $r = 0.05$, $p = 0.596$ | $r = 0.10$, $p = 0.336$ | **$r = 0.33$, $p < 0.01$** |
| Excitement | $r = 0.07$, $p = 0.474$ | $r = 0.19$, $p = 0.06$ | $r = 0.24$, $p = 0.02$ |
| Disappointment | $r = -0.02$, $p = 0.857$ | $r = 0.17$, $p = 0.093$ | $r = 0.05$, $p = 0.641$ |
| Irritation | $r = 0.11$, $p = 0.290$ | $r = 0.07$, $p = 0.476$ | $r = 0.18$, $p = 0.08$ |
| Delight | $r = 0.13$, $p = 0.183$ | $r = 0.02$, $p = 0.838$ | $r = 0.05$, $p = 0.638$ |
| Anxiety | $r = 0.13$, $p = 0.202$ | **$r = 0.27$, $p < 0.01$** | **$r = 0.36$, $p < 0.01$** |
| Pressure | $r = -0.03$, $p = 0.795$ | $r = 0.01$, $p = 0.951$ | $r = -0.01$, $p = 0.919$ |

[a]$n = 100$

The findings of the study provide theoretical and practical importance. From the theoretical point of view, it provided empirical evidence that gamers could feel positive as well as negative emotions. It also confirms that emotions could be dependent on one another. This supported the emotion theories of Cacioppo et al. [10], and Watson and Tellegen [3]. In terms of its practical value, it could serve as reminders for parents, game developers, and gamers. Parents are advised to monitor their children's gaming behavior. Gamer developers are encouraged to design games that could detect gamers' excessive gaming behaviors. A recommender system that enables game pause is a highly desired game design feature. Lastly, gamers themselves have to be cautious about their gaming behaviors and practice self-restraint. Hence, prudence and temperance are two virtues that gamers should learn to apply in these activities.

However, there are still research gaps that need to be addressed. First, there are other emotions, such as anger, excitement, frustration, boredom, and amusement that were not included in the present study. Second, it is also unclear as to what coping mechanisms gamers employed when they are confronted with game defeat. Lastly, it is also worth investigating the players' gaming behaviors after a game defeat. An investigation whether they continue the game or shift to another game and the explanation of such behavior could be initiated.

## 5. CONCLUSIONS AND RECOMMENDATIONS

Based on the findings presented, the null hypothesis stating that there is no significant relationship between emotions exhibited while playing computer games and the gaming behaviors of the respondents is partially rejected. Hence, not all gaming behaviors were related to the emotions exhibited while playing computer games. It also concluded that anxiety was a more sensitive emotion than stress. The study also revealed that there were two types of gamers. The first type of gamers was the prudent gamers who visited cyber cafés for no more than twice a week. On one hand, the second type were those who visited cyber excessively (i.e., at least three times a week).

The study also provided empirical evidence on Cacioppo et al. [10], and Watson and Tellegen [3] emotion theories. It was shown that independent and mixed emotions were shown while playing computer games. Additionally, these emotions could positive or negative.

The study calls for further investigation of other emotions that were not captured in the study. Emotions such as anger, excitement, boredom, frustration, and amusement can be investigated. It is also not yet known what coping mechanisms gamers may use and what their associated behaviors are when they are defeated in a game. Further study can shed light on this research gap.